\documentclass[preprint,eqsecnum,aps,epsfig,nofootinbib]{revtex4}
\usepackage{epsfig}

\newcommand{\be}{\begin{equation}}
\newcommand{\ee}{\end{equation}}
\newcommand{\bea}{\begin{eqnarray}}
\newcommand{\eea}{\end{eqnarray}}

\def\lap{\mathrel{\mathpalette\fun <}}

\def\fun#1#2{\lower3.6pt\vbox{\baselineskip0pt\lineskip.9pt
  \ialign{$\mathsurround=0pt#1\hfil##\hfil$\crcr#2\crcr\sim\crcr}}}

\def\lsim{\mathrel{\raise.3ex\hbox{$<$\kern-.75em\lower1ex\hbox{$\sim$}}}}
\def\gsim{\mathrel{\raise.3ex\hbox{$>$\kern-.75em\lower1ex\hbox{$\sim$}}}}
  
\begin{document}

\preprint{TKYNT-07-04, UT-07-08}

\title{Stau-catalyzed $^6$Li Production in Big-Bang Nucleosynthesis
\\ \ }
\author{
K. Hamaguchi$^{1}$, T. Hatsuda$^{1}$,  M. Kamimura$^{2}$,
Y. Kino$^{3}$ and T. T. Yanagida$^{1}$
\\ \  }
\vspace{1cm}
\affiliation{
$^{1}$ Department of Physics, The University of Tokyo, Tokyo 113-0033, Japan\\
$^{2}$ Department of Physics, Kyushu University, Fukuoka 812-8581, Japan\\
$^{3}$ Department of Chemistry, Tohoku University, Sendai 980-8578, Japan
}
\begin{abstract}
  If the gravitino mass is in the region from a few GeV to a few 10's GeV,
the scalar lepton $X$ such as stau is most likely the next lightest
 supersymmetry particle.
 The negatively charged and long-lived 
 $X^-$ may form a Coulomb bound state $(A X^-)$  with 
 a nucleus $A$ and may affect the big-bang nucleosynthesis
  through catalyzed fusion process.
 We calculate a production cross section
 of $^6{\rm Li}$ from the catalyzed fusion 
 $( ^4{\rm He} X^-) + d \rightarrow \, ^6{\rm Li} + X^-$ 
 by solving the Schr\"{o}dinger equation exactly for three-body system of
 $^4{\rm He}$, $d$ and $X$. We utilize the state-of-the-art coupled-channel
 method, which is known to be very accurate to describe 
 other three-body systems
 in nuclear and atomic reactions. The importance of the use of 
 appropriate nuclear potential and the exact treatment of the 
 quantum tunneling in the fusion process are emphasized.
  We find that the astrophysical $S$-factor at the Gamow peak 
  corresponding to
  $T=10$ keV is 0.038 MeV barn. This leads to
  the $^6{\rm Li}$ abundance from the catalyzed process as
   $^6 {\rm Li}|_{\rm CBBN}\simeq 4.3\times 10^{-11} 
 ({\rm D}/2.8\times 10^{-5})([n_{_{X^-}}/s]/10^{-16})$ 
 in the limit of long lifetime of $X$.
    Particle physics implication of this result is also
 discussed.
\end{abstract}

\maketitle

\section{Introduction}

The gravitino is the most important prediction of supergravity (SUGRA) \cite{SGRA}.
Its mass is expected to be in a wide range, $1$ eV$-100$ TeV, depending on the mediation
mechanism of supersymmetry (SUSY) breaking effects. In particular,
 the gravitino with a mass of $O(1)$ MeV$-O(10)$ GeV  is 
 very interesting in the sense that it is most likely the stable and  
 lightest  SUSY particle (LSP) and can be a candidate of cold dark matter in the universe.
  Moreover,  if the gravitino mass lies between a few GeV and a few 10's GeV
 and the next LSP (NLSP) is a charged scalar lepton ($X$) such as 
  stau ($\tilde{\tau}$), the SUGRA can be tested in high-energy
 collider experiments \cite{BHRY}. 
 The key observation is that the gravitino mass $m_{3/2}$,  
 the stau mass $m_{_X}$ and the stau lifetime $\tau_{_X}$ could be measured 
 in high precision if $m_{3/2}$ is of $O(10)$ GeV  (cf.~\cite{stauLHCILC}).
 Then the Planck scale $M_{\rm PL}$ is extracted 
 through the relation, $\tau_{_X}^{\rm (exp.)}
 =\tau_{_X}(m_{_X}^{\rm (exp.)}, m_{3/2}^{\rm (exp.)}, M_{\rm PL})$.
 This independent determination of $M_{\rm PL}$ will
  provide us with a crucial test of the SUGRA.

 For the gravitino mass in such an interesting region as discussed above,
 the stau has necessarily a long lifetime. For example, it is  of order one month
 for  $m_{3/2}\simeq 10$ GeV and $m_{X}\simeq 100$ GeV.
 However, such a  long-lived particle is potentially
  dangerous in cosmology. If it decays after the 
 big-bang nucleosynthesis (BBN), the decay products destroy the light elements 
 and ruin the success of the SBBN (standard BBN) \cite{Weinberg, KKM}. 
 More seriously, if the long-lived particle is negatively charged as $X^-$, 
 it forms bound states together with positively charged nuclei, which  
leads to an enhancement of some nuclear reaction rates.
 That is, $X^-$ plays as a catalyzer of nuclear fusion (catalyzed BBN or
  CBBN in short) and  affects cosmological 
  nuclear abundances.\footnote{Massive and long-lived $X^-$ produced
   in accelerators may be of practical use as a catalyzer for D-D and D-T fusions
   similar to the muon catalyzed fusion \cite{HHY}.}

 In a recent article  \cite{Pospelov},
  Pospelov has argued that too much $^6{\rm Li}$ is produced through 
  the CBBN if 
  the lifetime of $X^-$ is long enough,  $\tau_{_X} >  10^3-10^4$ sec,
  and its abundance is large enough, $n_{_{X^-}}/s > 10^{-17}$, where $n_{_{X^-}}$
   and $s$ are the number density of $X^-$ and the entropy density
   of the universe, 
     respectively.\footnote{Other effects of the $X^-$ 
     bound states have been considered in \cite{KT,KR}.}
  The production rate of $^6{\rm Li}$ estimated in Ref.~\cite{Pospelov} 
  is based on a 
  naive comparison of the standard process 
  $ ^4{\rm He}  + d \rightarrow\  ^6{\rm Li} + \gamma$ and the 
  new catalyzed process $( ^4{\rm He} X^-) + d \rightarrow\ ^6{\rm Li}+ X^-$
   where  $( ^4{\rm He} X^-)$ is the 1s Coulomb bound state of $^4{\rm He}$
   and $X^-$.  It is assumed  that the standard process is dominated
    by the E2 transition\footnote{This particular assumption itself 
    may not be justified since E1 transition could be comparable
    or even larger than E2 at low energies. See e.g. \cite{NWS01}.}
   induced by the interaction $Q_{ij} \nabla_i E_j$ with $Q_{ij}$ being
    the quadrapole operator.  Then, by
   applying the same interaction to the photonless process
   with $E_j$ replaced by the Coulomb field associated with $X^-$,
   it was concluded that the astrophysical $S$-factor of the new process
   at zero incident energy is about  0.3 MeV barn.
   However, the assumptions adopted in the above estimate
   do not have firm ground:
   For example, the angular momentum of the initial $^4{\rm He}$-$d$ system in 
    the catalyzed process is dominated by zero, $L=0$, while
    the angular momentum of the initial system in the standard process 
    is dominated by $L=2$ $(L=1)$ for the E2 (E1) transition.
   Such a kinematical difference invalidates the use of 
     the SBBN process to estimate the CBBN process.
   Furthermore, the quantum interplay between the $^4{\rm He}-d$ nuclear fusion and the 
   tunneling of $d$ through the Coulomb barrier plays
   an important role in low-energy catalyzed fusion and 
  cannot be treated in a perturbative manner. 
   
  The purpose of this paper is to solve the Shr\"{o}dinger equation 
  for the three-body system ($ ^4{\rm He}$, $d$ and $X^-$) 
 exactly and derive a reliable $S$-factor for the $X^-$ CBBN.
  The method we adopt is the state-of-the-art coupled-channel technique
   developed by two of the present authors (M.K. and Y.K.) 
 together with E. Hiyama 
  \cite{Kamimura1988b,Kamimura1988,Kino1993,Hiyama2003}. 
  Since we can treat the catalyzed process directly, we do not need to
   refer to any of the SBBN processes.
  Also, the method has been already proven to be highly accurate
  and useful in atomic and  nuclear physics 
 (reviewed in \cite{Hiyama2003}).   
 We find that the obtained astrophysical $S$-factor at 
  $E=36.4$ keV (the position of the Gamow peak for $T=10$ keV which 
   is relevant to the CBBN) is 0.038 MeV~barn. This is 
about 10 times smaller than the estimate in Ref.~\cite{Pospelov} 
at the same energy.
For long lived  stau, our $S$-factor leads to the $^6 {\rm Li}$ abundance from CBBN as $^6 {\rm Li}|_{\rm CBBN}\simeq 4.3\times 10^{-11}({\rm D}/2.8\times 10^{-5})([n_{_{X^-}}/s]/10^{-16})$.
Therefore, an observational upper bound on the $^6 {\rm Li}$ abundance 
$^6 {\rm Li}<6.1\times 10^{-11}$ ($2\sigma$)~\cite{KKM} leads to a bound on the $X^-$ abundance, $n_{X^-}/s<1.4\times 10^{-16}$.
This requires a dilution of the relic stau by some entropy production at late time by a factor of $\Delta\simeq (300-600)\times (m_{\widetilde{\tau}} / 100~\mathrm{GeV})$. 
Such a dilution factor can easily be consistent with the thermal leptogenesis~\cite{FY} for a reheating temperature $T_R\gsim 10^{12}$ GeV.

 The organization of this paper is as follows.
 In Sec.II, we summarize the basic reaction of the $X^-$ catalyzed fusion
  as well as its atomic analogue, the muon transfer reaction.
 In Sec.III, we present our method of solving the 
 $^4{\rm He}$-$d$-X$^-$  three-body problem.
 In Sec.IV, we show our result of astrophysical $S$-factor as a function
  of the incident energy. 
   Sec.V is devoted to conclusion and discussions in which
  the $^6{\rm Li}$ abundance in CBBN and its 
   implication to particle physics are also mentioned.

\section{The $X^-$ catalyzed process }
 
 As we have mentioned in the Introduction, we will treat the 
 following process directly in a fully quantum mechanical manner:
\begin{equation}
( ^4{\rm He} X^-) + d \rightarrow \, ^6{\rm Li} + X^- + 1.1 
\,{\rm MeV} .
\label{eq:transfer}
\end{equation}
As shown later in Section V,
 the relevant temperature $T$ for CBBN is about 10 keV.
  This corresponds to the Gamow peak of the incident channel located 
   at  the c.m. energy ${E}_{_{\rm G}}=$36.4 keV with the full  $1/e$ width  of 44 keV.
 Thus the c.m. energy of the incident channel
  will be typically up to  $100$ keV. 
We adopt the $^4{\rm He}-d$ cluster model in which
 the $^6$Li nucleus is a bound state of a $^4$He nucleus and a deuteron. 
This model has extensively been utilized in the studies of the structure and reactions of 
light nuclei and is very well established
 (see e.g. \cite{Ikeda1980,Kamimura1986}).
Such a cluster-model treatment  allows us to investigate the reaction (\ref{eq:transfer}) 
 as a quantum three-body problem, $ ^4{\rm He}+d+X^-$.  The rearrangement
 of the three ingredients takes place during the 
 reaction process through the long range Coulomb interaction and 
the short range nuclear interaction.

 Although solving the three-body problem accurately  is quite an elaborate task,
two of the authors (M.K. and Y.K.) have an experience 
\cite{Kamimura1988,Kino1993,Hiyama2003} 
to have solved a similar subject, the muon transfer reaction,
\begin{equation}
(d\mu^-)_{1s}+t \rightarrow (t\mu^-)_{1s} +d + 48\ {\rm eV}
\label{eq:muon}
\end{equation}
at incident energies of $0.001-100$ eV
in the context of the muon catalyzed D-T fusion cycle
\cite{Nagamine1998}.
Among the calculations of this reaction in the literatures, 
their three-body coupled-channel method 
was found to provide with the most accurate result 
(see $\S$8 of \cite{Hiyama2003}).

In the present paper, the same coupled-channel method is applied to the reaction, Eq.
(\ref{eq:transfer}). We note here that
it is very important to 
employ an appropriate nuclear interaction between 
$^4$He and deuteron in treating the reaction; we choose the interaction
 which can reproduce the binding energy and the 
charge form-factor of $^6$Li as well as
the low-energy $^4{\rm He}-d$ scattering phase shift.


\section{Coupled channel method}

\begin{figure}[htb]
\begin{center}
\epsfig{file=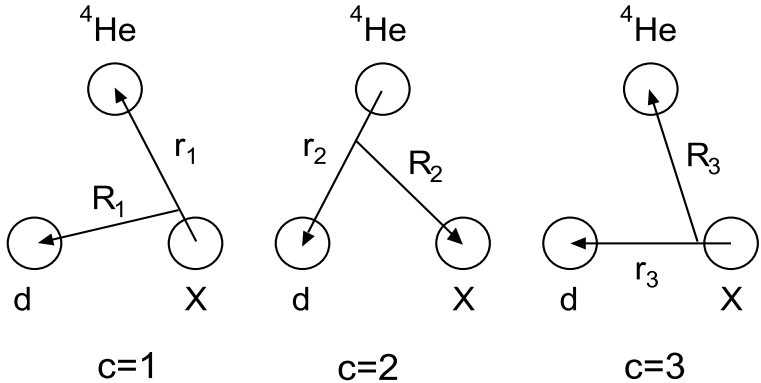,width=10cm,height=5cm}
\end{center}
\caption[]{
Three sets of Jacobi coordinates of
the $ ^4{\rm He}+d+X^-$ system. The entrance (exit)
 channel is described by the coordinate system of 
 $c=1$ ($c=2$). 
}
\label{fig:stau_fig1}
\end{figure}

 Following Ref.\cite{Hiyama2003}, we briefly
 explain the three-body coupled-channel method 
 to investigate the reaction (\ref{eq:transfer}).
 As shown in Fig.\ref{fig:stau_fig1}, we consider all the
 Jacobi coordinates $({\bf r}_c, {\bf R}_c)$
 of the three possible sets, $c=1,2$ and $3$.
 The entrance channel $( ^4{\rm He} X^-) + d $ 
 and the exit channel $^6{\rm Li} + X^-$ in our reaction
 are best described 
 by the coordinates for $c=1$ and those for $c=2$, respectively.
 Since we are interested in  the incident energy of the entrance channel 
 below 100 keV, there is no other open channel
 than the above two.  All the excited states of
 $( ^4{\rm He} X^-)$ and $^6$Li as well as all the states of
 $(dX^-)$ can be excited only virtually  in the intermediate stage of the
 reaction. To describe such an intermediate state 
 where particle-rearrangement takes place, it is convenient
 to utilize the coordinates for $c=3$ together with those
  for $c=1,2$.

\noindent
\subsection{Three-body Schr\"{o}dinger equation}

The Schr\"{o}dinger equation for  
 the total wave function $\Psi_{JM}$ 
of the $ ^4{\rm He}+d+X^-$ system
having an angular momentum $J$ and its $z$-component $M$
is given by 
\begin{equation}
 ( H - E_{\rm tot} ) \Psi_{JM} = 0 ,  
\label{eq:schroedinger}
\end{equation}
with the Hamiltonian,
\begin{eqnarray}
H=-{\hbar^2 \over 2m_{c}} \nabla^{2}_{{\bf r}_{c}}
           -{\hbar^2 \over 2M_{c}} \nabla^{2}_{{\bf R}_{c}}
          + V_{^4{\rm He}{\mbox -}X}(r_1) 
          + V_{^4{\rm He}{\mbox -}d}(r_2) 
          + V_{d{\mbox -}X}(r_3) . \qquad 
\label{eq:hamiltonian}
\end{eqnarray}
As far as we use the reduced masses
 ($m_c$ and $M_c$)  associated with the coordinates 
 (${\bf r}_c$ and ${\bf R}_c$),  every choice of $c$ in the 
  kinetic term is equivalent.
 $V_{A_c{\mbox -}B_c}(r_c)$  denotes
the potential between the particles $A_c$ and $B_c$
$(A_1{\mbox -}B_1\!=\!\,^4{\rm He}{\mbox -}X, 
A_2{\mbox -}B_2\!=\!\,^4{\rm He}{\mbox -}d, A_3{\mbox -}B_3= d{\mbox -}X)$
and will explicitly be given  below. 

Spin of the deuteron is neglected, and therefore the angular momentum
of the ground states of $^6$Li is  zero as well as that of
$( ^4{\rm He} X^-)$.
We denote the ground-state wave function
of  $( ^4{\rm He} X^-)$  in 
$c=1$ by $\phi_{\rm g.s.}^{(1)}({\bf r}_1)$
and its eigenenergy by $\varepsilon_{\rm g.s.}^{(1)}$,
and similarly, that of
$^6$Li in $c=2$ by $\phi_{\rm g.s.}^{(2)}({\bf r}_2)$
and its eigenenergy by $\varepsilon_{\rm g.s.}^{(2)}$.
They are obtained by solving
\begin{equation}
  \big[  -{\hbar^2 \over 2m_{c}} \nabla^{2}_{{\bf r}_{c}}
     +   V_{A_c{\mbox -}B_c}(r_c) - 
\varepsilon_{\rm g.s.}^{(c)} \,\big] \,
   \phi_{\rm g.s.}^{(c)}({\bf r}_c) =0 . \qquad (c=1,2) 
\label{eq:subsystem}
\end{equation}

Center-of-mass scattering energy of the channel $c$ 
associated with the coordinate ${\bf R}_c$, say $E_c$, 
is introduced by
$E_c= E_{\rm tot} - \varepsilon_{\rm g.s.}^{(c)}$ together with 
the corresponding wave number $k_c$ by $\hbar^2 k^2_c /2M_c=E_c \, (c=1,2)$. 
For a given total energy $E_{\rm tot}$, the Schr\"{ o}dinger equation 
(\ref{eq:schroedinger}) should be solved
under the scattering boundary condition:
\begin{eqnarray}
\lim_{R_{c}\to\infty}\Psi_{JM} &\!=\!&
\phi_{\rm g.s.}^{(c)}({\bf r}_c) 
 \Big[ u^{(-)}_{J}(k_{c}R_{c})\delta_{c \,1}
 -\sqrt{\frac{v_1}{v_{c}}}
S^J_{1 \to c}u^{(+)}_{J}(k_{c}R_{c}) \Big] Y_{JM}(\widehat{\bf R}_c),
\qquad (c=1,2) \qquad
\label{eq:boundary}
\end{eqnarray}
where
$u^{(\pm)}_{J}(kR)(=(G_J(kR) \pm i F_J(kR))/kR)$  
are the asymptotic outgoing and incident
Coulomb wave functions. 
$S^J_{1 \to c }$ is  the S-matrix for the transition
from  the channel 1 to $c$ and 
$v_c$ is the velocity of the channel $c$. 
 By introducing a simplified notation, $E\equiv E_1$ and $k\equiv k_1$,
the cross section of the rearrangement process 
(\ref{eq:transfer})
is given by 
\begin{equation}
 \sigma_{1 \to 2}(E) 
=\frac{\pi}{k^2}\sum_{J=0}^{\infty} (2J+1)
        \bigl|S^J_{1 \to 2} \bigr|^2  ,
\label{eq:cross}
\end{equation}
and the astrophysical $S$-factor is derived from 
\begin{equation}
S(E)= \sigma_{1 \to 2}(E)\,E \, {\rm exp}(2\pi \eta(E)),
\label{eq:astro}
\end{equation}
where $\eta(E)$ is the Sommerfeld parameter of the entrance
channel.

The three-body wave function which describes the transfer reaction
(\ref{eq:transfer}) and the elastic
$(^4{\rm He}X^-)+d$ scattering
simultaneously is written as
\begin{equation}
\Psi_{JM}=\phi^{(1)}_{\rm g.s.}({\bf r}_1)\,\chi^{(1)}_{JM}({\bf R}_1)
         +\phi^{(2)}_{\rm g.s.} ({\bf r}_2)\,\chi^{(2)}_{JM}({\bf R}_2)
         +\Psi^{({\rm closed})}_{JM}\;.
\label{eq:psi12}
\end{equation}
The first and the second terms represent 
the open channels, $c=1$ for $(^4{\rm He}X^-)+d$  and
 $c=2$ for $ ^6{\rm Li}+X^-$. The factors
$\chi^{(c)}_{JM}({\bf R}_c)
(= \chi^{(c)}_J(R_c) \, Y_{JM}(\widehat{\bf R}_c)\,)$ 
 describe the scattering waves along the coordinates ${\bf R}_c$
and are to be solved under the boundary condition
(\ref{eq:boundary}).
The third term, $\Psi^{({\rm closed})}_{JM}$,
stands for all the closed (virtually-excited) channels 
in the energy range of this work; 
in other words, this term is introduced to represent
all the asymptotically-vanishing three-body amplitudes  
that are not included in the first two scattering terms. 
For example, the third term describes such an effect 
that the incoming deuteron
attracts the $^4$He in the 1s-orbit around $X^-$ 
and distorts the orbit before picking up the $^4$He.

Since $\Psi^{({\rm closed})}_{JM}$
vanishes asymptotically, it is  reasonable
and useful  
\cite{Kamimura1988,Kino1993,Hiyama2003} 
to expand it in terms of a 
complete set of  $L^2$-integrable three-body basis functions, 
$\{ \Phi_{JM, \nu}; \nu=1-\nu_{\rm max} \}$,
spanned in a finite spatial region (see Sec.3.C):
\begin{equation}
\Psi^{({\rm closed})}_{JM}  =  \sum_{\nu=1}^{\nu_{\rm max}}
  b_{J\,\nu} \; \Phi_{JM,\:\nu}     .
\label{eq:Phi-def}
\end{equation}
Equations for $\chi^{(1)}_{J}(R_1)$, 
$\chi^{(2)}_{J}(R_2)$ 
and the coefficient $b_{J \nu}$ are given by the $\nu_{\rm max}+2$
simultaneous equations
\begin{equation}
\langle \, \phi^{(c)}_{\rm g.s.}({\bf r}_c)   Y_{JM}(\widehat{\bf R}_c)\,
  \:   | \,H - E_{\rm tot} \,|\:
\Psi_{JM} \,\rangle _{{\bf r}_c,\, \widehat{\bf R}_c }=0 ,
           \hskip 2 em (c=1,2) 
\label{eq:eq26}
\end{equation}
and
\begin{equation}
\langle \,\Phi_{JM,\,\nu}\:|\, H - E_{\rm tot} \, | \:\Psi_{JM} \,\rangle =0 .
    \hskip 20 pt  (\nu=1 - \nu_{\rm max})  
\label{eq:eq27}
\end{equation}
Here, $\langle \hskip 3 ex \rangle_{{\bf r}_c,\, \widehat{\bf R}_c}$
 denotes the integration over 
${\bf r}_{c}$ and $\widehat{\bf R}_c$.

Since $\Phi_{JM,\:\nu}$ are constructed 
so as to diagonalize the three-body Hamiltonian as
\begin{equation}
\langle \Phi_{JM,\,\nu}| \,  H \,  |\Phi_{JM,\,\nu'}\rangle
           =  E_{J\,\nu} \delta_{\nu \nu'} ,
  \qquad  (\nu,\nu'=1 - \nu_{\rm max}) 
\label{eq:diag}
\end{equation}
the coefficients $b_{J \nu}$ can be written, from Eqs.(\ref{eq:eq27}),
as
\begin{eqnarray}
b_{J \nu}=\frac{-1}{E_{J \nu}-E_{\rm tot}} 
 \langle \Phi_{JM,\, \nu}\,|\, H-E_{\rm tot} \,|\,
  \phi^{(1)}_{\rm g.s.}\chi^{(1)}_{JM}
 +\phi^{(2)}_{\rm g.s.}\chi^{(2)}_{JM}\rangle . \qquad
 (\nu=1 - \nu_{\rm max}) \qquad
\label{eq:bjnu}
\end{eqnarray}
Inserting Eqs.(\ref{eq:bjnu}) into $b_{J \nu}$ in $\Psi_{JM}$
in Eqs.(\ref{eq:eq26}), 
we reach two coupled integro-differential
equations for $\chi^{(1)}_{J}(R_1)$ 
and $\chi^{(2)}_{J}(R_2)$. 

 The integro-differential equations, though 
not recapitulated here 
 (c.f. \S 8 of \cite{Hiyama2003} for them),
are solved by
using both the direct numerical method (finite-difference method)
and the Kohn-type variational method
\cite{Kamimura1977,Hiyama2003},  and we have obtained
the same result for the incident energies relevant 
 to CBBN, $E>10$ keV.  
 The coupling between the entrance and exit channels as well as 
  the contribution from the closed channels 
 $\Psi_{JM}^{\rm (closed)}$  were found to be significantly large
  as will be discussed later.

\noindent
\subsection{Nuclear potentials}

It is essential for the three-body calculation to
employ appropriate nuclear interaction between 
$^4$He and deuteron which governs the $^4$He-transfer process.
In this subsection, we define the potentials 
$V_{^4{\rm He}{\mbox -}X}(r_1),V_{^4{\rm He}{\mbox -}d}(r_2)$ 
and $V_{d{\mbox -}X}(r_3)$ in our  Hamiltonian (\ref{eq:hamiltonian}).
First, we assume Gaussian-shape charge distributions of
$^4$He and deuteron as 
$2e(\pi b_1^2)^{-3/2} {\rm e}^{-(r/b_1)^2}$ and
$e(\pi b_3^2)^{-3/2} {\rm e}^{-(r/b_3)^2}$, respectively.  
 We take $b_1=1.37$ fm 
and $b_3=1.75$ fm, which reproduce 
observed r.m.s. charge radii,
$1.68$ fm  of $^4$He   and 
2.14 fm of deuteron. 
The potential between $^4$He and $X^-$ is then given by 
\begin{equation}
V_{^4{\rm He}{\mbox -}X}(r_1)= -\, 2\, e^2 \, \frac{{\rm erf}(r_1/b_1)}{r_1} ,
\end{equation}
and that between deuteron and $X^-$ is written as 
\begin{equation}
V_{d{\mbox -}X}(r_3)= - \, e^2 \, \frac{{\rm erf}(r_3/b_3)}{r_3}, 
\end{equation}
where ${\rm erf}(x)=\frac{2}{\sqrt{\pi}} \int_0^x e^{-t^2} {\rm d}t$ 
is the error function.
Energy of the $(^4{\rm He}X^-)_{1s}$ state is 
$ \varepsilon_{\rm g.s.}^{(1)}=-337.33$ keV
and the r.m.s. radius is 
$ \langle \,  r_1^2 \,\rangle^{1/2} =6.84$ fm.

The potential $V_{^4{\rm He}{\mbox -}d}(r_2)$ 
is a sum of the nuclear potential,
$V_{^4{\rm He}{\mbox -}d}^{\rm N}(r_2)$, 
and the Coulomb potential, $V_{^4{\rm He}{\mbox -}d}^{\rm C}(r_2)$.
The latter is given by
\begin{equation}
V_{^4{\rm He}{\mbox -}d}^{\rm C}(r_2)= 2 \, e^2
 \, \frac{{\rm erf}(r_2/\sqrt{b_1^2+b_3^2})}{r_2} . 
\end{equation}
The nuclear potential is assumed to have a two-range
Gaussian shape as
\begin{equation}
V_{^4{\rm He}{\mbox -}d}^{\rm N}(r_2) = v_0 \,e^{-(r_2/a)^2} + v'_0 \,e^{-(r_2/a')^2} 
\end{equation}
with $a=0.9$ fm, $v_0=500.0$ MeV, $a'=2.0$ fm and $v'_0=-64.06$ MeV.
The first term, a repulsive core, is introduced to
simulate the Pauli exclusion principle that
nucleons in the incoming 
deuteron should not occupy the nucleon s-orbit in
the $^4$He nucleus during the reaction process
(see e.g. \cite{Ikeda1980} for this role of
Pauli principle).
\begin{figure}[htb]
\begin{center}
\epsfig{file=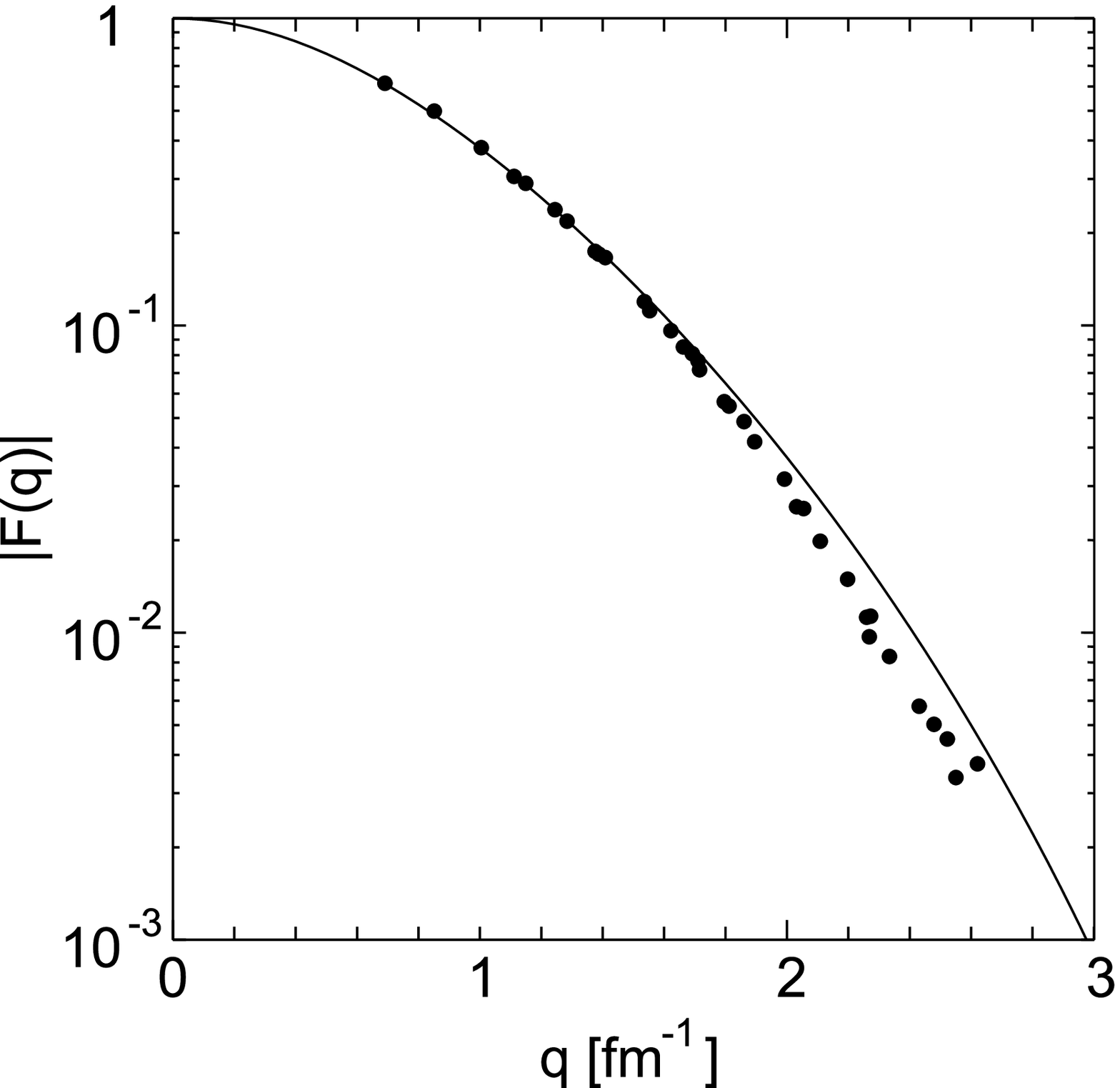,width=6.4cm}
\hspace{0.5cm}
\epsfig{file=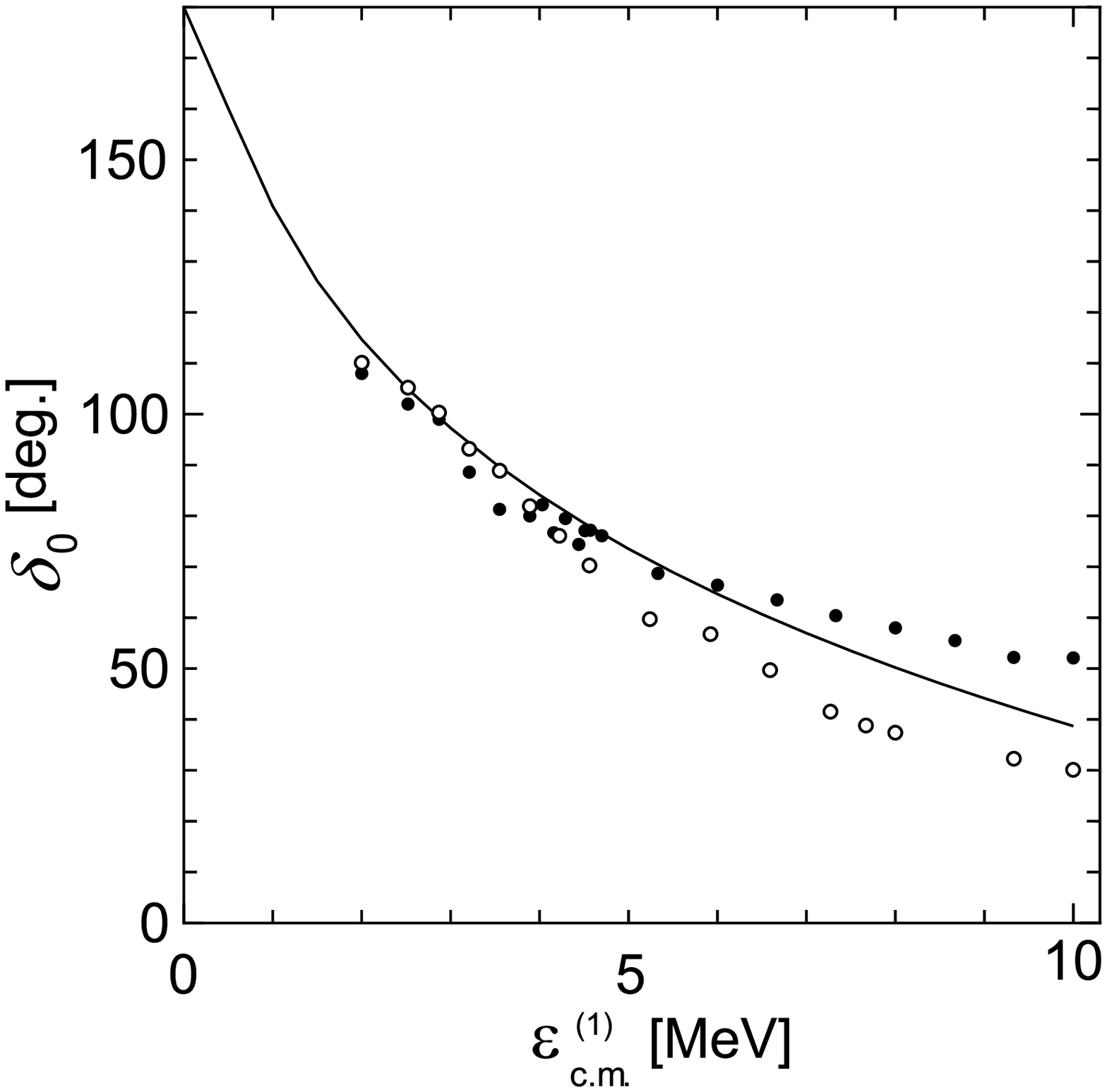,width=6.6cm}
\end{center}
\caption[]{Left panel:
Charge form factor of the electron scattering
from $^6$Li. The calculated values (experimental data \cite{Suelzle1967})
 are shown by the solid line (filled circles).
Right panel: The s-wave phase shift $\delta_0$ of the $^4{\rm He}+d$ scattering 
at c.m. energy $\varepsilon_{\rm c.m.}^{(1)} < 10 $ MeV.
 The calculated values are shown by the solid line, while
  the data from the phase-shift analysis are shown 
 by filled circles  \cite{Jenny1983} and  by open circles  \cite{Gruebler1975}.
}
\label{fig:stau_fig2}
\end{figure}

\begin{figure}[htb]
\begin{center}
\epsfig{file=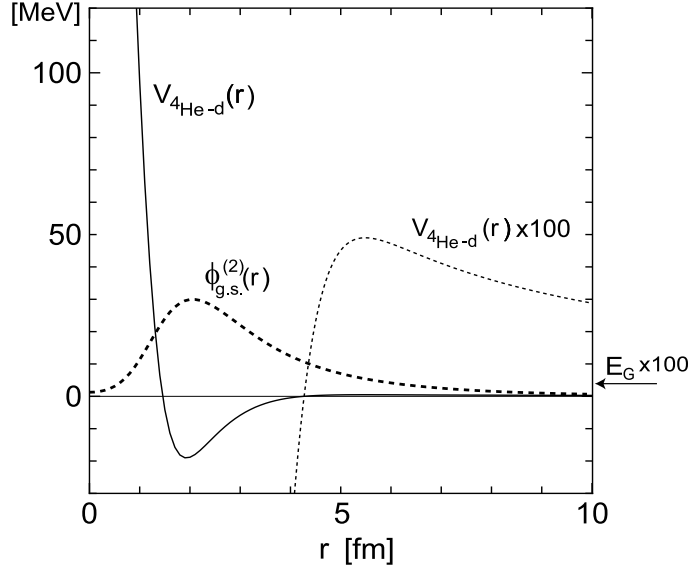,width=9cm,height=7.5cm}
\end{center}
\caption[]{
 The potential $V_{^4{\rm He}{\mbox -}d}(r)$ between $^4{\rm He}$ and deuteron
 (the solid line). To show its Coulomb barrier tail,  the same potential scaled 
  by 100 is shown by the dotted line together with the 
 typical incident kinetic energy $E_{\rm G}\times 100$ denoted by the arrow.  
  The dashed line shows the $^4{\rm He}-d$ relative 
 wave function in the $^6{\rm Li}$ ground state in an arbitrary unit,
 $\phi_{\rm g.s.}^{(2)}(r)$.}
\label{fig:stau_fig4}
\end{figure}

The above parameters of $V_{^4{\rm He}{\mbox -}d}^{\rm N}(r_2)$ 
were so determined that
the solution to the Schr\"{o}dinger equation 
(\ref{eq:subsystem}) could reproduce 
observed values of the energy $\varepsilon_{\rm g.s.}^{(2)}=
-1.474$ MeV and the r.m.s. charge radius 2.54 $\, {\rm fm}$
of the ground state of $^6$Li. 
 Furthermore,  the charge density of $^6$Li 
reproduces observed charge form factor 
of the electron scattering from $^6{\rm Li}$ as shown in the left panel of
 Fig.\ref{fig:stau_fig2}.
Simultaneously, use of the potential $V_{^4{\rm He}{\mbox -}d}(r_2)$ 
explains the low-energy s-wave phase shifts 
of the $^4{\rm He}+d$ scattering as shown in the right panel of
Fig.\ref{fig:stau_fig2}.
 Here, it should be noted that simple attractive potential without a repulsive 
 core leads to the phase shift which increases with increasing $E$
 and cannot explain the observed data.
 We thus have a good $^4{\rm He}-d$  potential $V_{^4{\rm He}{\mbox -}d}(r_2)$
 in order to perform a precise study
 of the three-body reaction (\ref{eq:transfer}).
 The potential $V_{^4{\rm He}{\mbox -}d}(r_2)$ 
 and the wave function $\phi^{(2)}_{\rm g.s.}(r_2)$
 are illustrated in Fig.\ref{fig:stau_fig4} in which $r_2$ is denoted
  by $r$ for simplicity.

\subsection{Three-body basis functions}

The $L^2$-integrable three-body basis functions
$\{ \Phi_{JM, \nu}; \nu=1-\nu_{\rm max} \}$
used in (\ref{eq:Phi-def}) to expand 
$\Psi^{({\rm closed})}_{JM}$ are introduced as follows
\cite{Hiyama2003}:
$\Phi_{JM, \,\nu}$ are written as a sum of the component
functions in the Jacobi-coordinate sets $c=1-3$ (Fig.\ref{fig:stau_fig1}),
\begin{equation}
\Phi_{JM, \,\nu}=\Phi_{JM, \,\nu}^{(1)}({\bf r}_1, {\bf R}_1)
+\Phi_{JM, \,\nu}^{(2)}({\bf r}_2, {\bf R}_2)
+\Phi_{JM, \,\nu}^{(3)}({\bf r}_3, {\bf R}_3) \; .
\end{equation}
Each component  is expanded in terms of the Gaussian basis functions
of the coordinates ${\bf r}_c$ and ${\bf R}_c$:
\begin{equation}
\Phi_{JM, \,\nu}^{(c)}({\bf r}_c, {\bf R}_c)  
=\sum_{n_cl_c,N_cL_c}A^{(c)}_{J \,\nu,\, n_cl_c,\,N_cL_c} \:
\left[\phi^{\rm G}_{n_cl_c}({\bf r}_c)\:
\psi^{\rm G}_{N_cL_c}({\bf R}_c)
\right]_{JM}
\qquad (c=1-3)\;,
\label{eq:bases}
\end{equation}
where 
\begin{eqnarray}
&&\phi^{\rm G}_{nlm}({\bf r}) =
r^l\:e^{-(r/r_n)^2} \:Y_{lm}({\widehat {\bf r}}) , \quad
\quad \qquad  r_n=r_1\, a^{n-1} , \; \qquad (n=1-n_{\rm max}) 
\label{eq:3gaussa}\\
&&\psi^{\rm G}_{NLM}({\bf R}) = 
R^L\:e^{-(R/R_N)^2} \:Y_{LM}({\widehat {\bf R}}) , \quad
R_N=R_1\, A^{N-1} , \;\quad (N=1-N_{\rm max})
\label{eq:3gauss}
\end{eqnarray}
The Gaussian ranges are postulated to lie in
geometric progression.
Basis functions so chosen are suited for describing
both short range correlations (mainly due to the
$^4{\rm He}-d$ nuclear interaction) 
and long  range asymptotic behavior
simultaneously
\cite{Hiyama2003}, and therefore 
they are efficient to describe the three-body configurations
(closed-channel contribution) in the interaction region
in the intermediate stage of reactions 
\cite{Kamimura1988,Kino1993,Hiyama2003}. 
The coefficients 
$A^{(c)}_{J \,\nu,\, n_cl_c,\,N_cL_c}$
in (\ref{eq:bases})
are determined by diagonalizing the three-body Hamiltonian $H$
as Eqs.(\ref{eq:diag}).
   
   In the calculation for $J=0$, we took $l_c=L_c=0, 1, 2$ and
$n_{\rm max}=N_{\rm max}=15$ for $c=1-3$.  Therefore, 
total number of the three-body Gaussian basis functions
$\left[\phi^{\rm G}_{n_cl_c}({\bf r}_c)\:
\psi^{\rm G}_{N_cL_c}({\bf R}_c)
\right]_{JM}$
to construct $\{\Phi_{JM, \,\nu}\}$
amounts  to $\nu_{\rm max}=2025$, 
which was found to be large enough for the
present calculation.  As for the Gaussian ranges,
we took $r_1, r_{n_{\rm max}}, R_1$ and $R_{N_{\rm max}}$ 
to be 0.5, 15.0, 1.0, 40.0 fm, which are sufficiently precise
for the present purpose. The expansion
(\ref{eq:Phi-def}) was found to converge quickly 
with  increasing $\nu$, and $\nu \lap 100 
\; (E_{J \nu} \lap 1$ MeV above 
the $(^4{\rm He}X^-)-d$ threshold) is very sufficient. 

\section{Result of the reaction rate}

\begin{figure}[bth]
\begin{center}
\epsfig{file=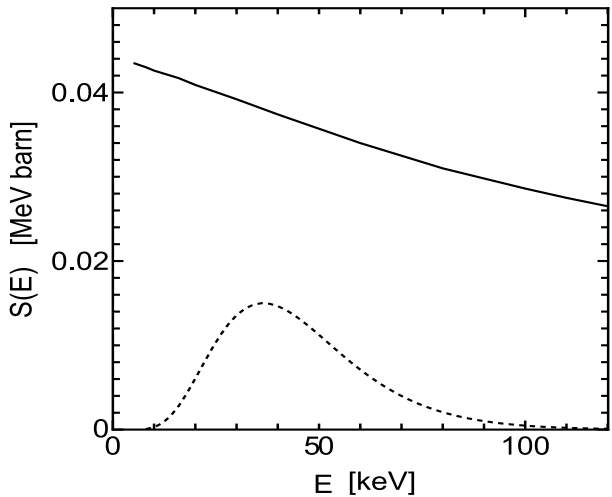,width=8cm,height=6.5cm}
\end{center}
\caption[]{The 
astrophysical $S$-factor (the solid line) obtained  by the
full coupled-channel calculation. The dashed line illustrates
the Gamow peak (in arbitrary units)
for $T=10$ keV with the peak maximum at $E_{\rm G}=36.4$ keV.
}
\label{fig:stau_fig5}
\end{figure}

Table \ref{table:percent} lists  the  cross section $\sigma_{1 \to 2}(E)$
and the astrophysical $S$-factor $S(E)$
of the reaction (\ref{eq:transfer})
obtained  by the full coupled-channel calculation 
at several energies around  $E_{\rm G}=36.4 $ keV.
In Fig.\ref{fig:stau_fig5},
$S(E)$ is shown  together with the Gamow peak.
The $S$-factor increases with decreasing energy $E$ since
there are three-body  bound states below the $^6{\rm Li}$-$X^-$ threshold.
 Note that  $S(E_{\rm G})$ estimated in  \cite{Pospelov}
is an order of magnitude larger than the value
given by the present exact calculation.

Since the $S$-factor can be approximated around $E_{\rm G}$ as 
\begin{equation}
S(E) \simeq S(E_{\rm G}) + a\,(E - E_{\rm G} ),
\label{eq:expansion}
\end{equation}
with $a=\left( {\partial S}/{\partial E}\right)_{E_{\rm G}}$, 
the reaction rate, using Eqs.(4.56) and (4.74) of 
\cite{Clayton1983}, is expressed as
\begin{equation}
N_A \,\langle \, \sigma v \,\rangle = 
6.24  \times 10^9 \,S(E_{\rm G})\,
  (1 - 0.0718 \,a \,T_9/S(E_{\rm G}))\, T_9^{-2/3} \, 
{\rm exp}\,(- 5.33 \,T_9^{-1/3}) \;\; 
{\rm cm}^3 \,{\rm s}^{-1} \, {\rm mol}^{-1},
\end{equation}
where $N_A$ is the Avogadro constant and 
$S(E_{\rm G})$, $a$ and $T_9$  are to be given in units 
of MeV$\!$~barn, barn and $10^9$ K,
respectively.
Taking $S(E_{\rm G})=0.0380$ MeV barn and $a=-0.18$ barn, we obtain
\begin{equation}
N_A \,\langle \, \sigma v \,\rangle = 
2.37  \times 10^8 \,(1 - 0.34 \,T_9)\, T_9^{-2/3} \, 
{\rm exp}\,(- 5.33 \,T_9^{-1/3}) \quad 
{\rm cm}^3 \,{\rm s}^{-1} \, {\rm mol}^{-1}.
\label{eq:sigmav}
\end{equation}

%
\begin{table}[th]
\caption{The cross section $\sigma_{1 \to 2}(E)$
and the astrophysical $S$-factor $S(E)$ 
of the reaction $(^4{\rm He} X^-) + d \rightarrow \, ^6{\rm Li}+ X^-$
obtained by the full coupled-channel calculation. 
The most effective  energy $E_{\rm G}$ for $T=$10 keV is 36.4 keV.
}
\label{table:1}
\begin{center}
\begin{tabular}{cccc} 
\hline \hline
\noalign{\vskip 0.2 true cm} 
 $E$   [keV]   &  $\quad$   $\sigma_{1 \to 2}\,$ [barn]  &
 $\quad$  $S\,$ [MeV barn]  & $\quad$  \\
\noalign{\vskip 0.2 true cm} 
\hline 
\vspace{-5 mm} \\
 10 & $\quad$   $3.85 \times 10^{-6}$  & 0.0426  \\
 20 & $\quad$   $1.09 \times 10^{-4} $  & 0.0410  \\
36.4& $\quad$   $6.88 \times 10^{-4}$  & 0.0380 \\
 50 & $\quad$   $1.41 \times 10^{-3}$  & 0.0357 \\
 100& $\quad$   $3.50 \times 10^{-3}$  & 0.0286 \\
\vspace{-5 mm} \\
\noalign{\vskip 0.1 true cm} 
\hline 
\hline\\
\end{tabular}
\label{table:percent}
\end{center}
\end{table}
%

  To check the necessity of the exact coupled-channel treatment  
  (\ref{eq:psi12})$-$(\ref{eq:bjnu}),
  we have calculated the cross sections in the following 
  approximations: 
  (i)  A coupled-channel 
  treatment with only the entrance and exit channels where
   the closed channel $\Psi_{JM}^{\rm (closed)}$ in Eq.(\ref{eq:psi12})
  is neglected, 
  and (ii)  an one-step approximation where
  the transition from the entrance channel to the exit channel in 
  the case (i)
   is treated in the leading order perturbation.
  For the case (i), the cross section $\sigma_{1 \to 2}$ 
  become approximately 3 times smaller than 
 those in Table \ref{table:percent}. For the 
  case (ii), the  cross section becomes an order of magnitude
 larger than that in the case (i). Therefore, these
  approximations are not justified.

 We have also checked that the cross section $\sigma_{1 \to 2}$
 is dominated by the total orbital angular momentum $J=0$, and
 contribution from the higher angular momenta $J \geq 1$ is
 three orders of magnitude smaller. 
 Furthermore, any pair among $^4$He, $d$ and $X^-$ 
  has dominantly zero orbital angular momentum, i.e. s-wave. This is in 
  sharp contrast to the SBBN reaction $^4{\rm He} + d \to \,^6{\rm Li} + \gamma $
  where the entrance channel is dominated either by d-wave (the E2 transition)
  or by p-wave (the E1 transition) from the selection rule.

\section{Conclusions and discussion}

Since we consider the gravitino masses of order 10 GeV, the scalar lepton $X$
(such as stau)  is most likely the next lightest
 supersymmetry particle and could have a long lifetime more than $10^3$ sec.
Thus, the negatively charged $X^-$ forms a bound states $(A X^-)$  with positively
 charged nuclei $A$ during the big-bang nucleosynthesis (BBN). Then, 
 some nuclear abundances are enhanced  through the catalyzed process.
 
 In this Letter, we have calculated a production cross section
 of $^6{\rm Li}$ from the catalyzed process 
 $( ^4{\rm He} X^-) + d \rightarrow \, ^6{\rm Li} + X^-$ 
 by exactly solving the Schr\"{o}dinger equation for three-body system of
 $^4{\rm He}$, $d$ and $X$. We utilize the state-of-the-art coupled-channel
 method, which is known to be very accurate to describe other three-body systems
 in nuclear and atomic reactions. The use of 
 appropriate nuclear potential and the exact treatment of the 
 quantum tunneling in the fusion process turned out to be important.
  We have found that the astrophysical $S$-factor at the Gamow peak corresponding to
  $T=10$ keV is 0.038 MeV barn as shown in 
  Table ~\ref{table:1}, Fig.~\ref{fig:stau_fig5} and Eq.~(\ref{eq:sigmav}).
 They are the main results of this Letter.

\begin{figure}[t]
\begin{center}
\epsfig{file=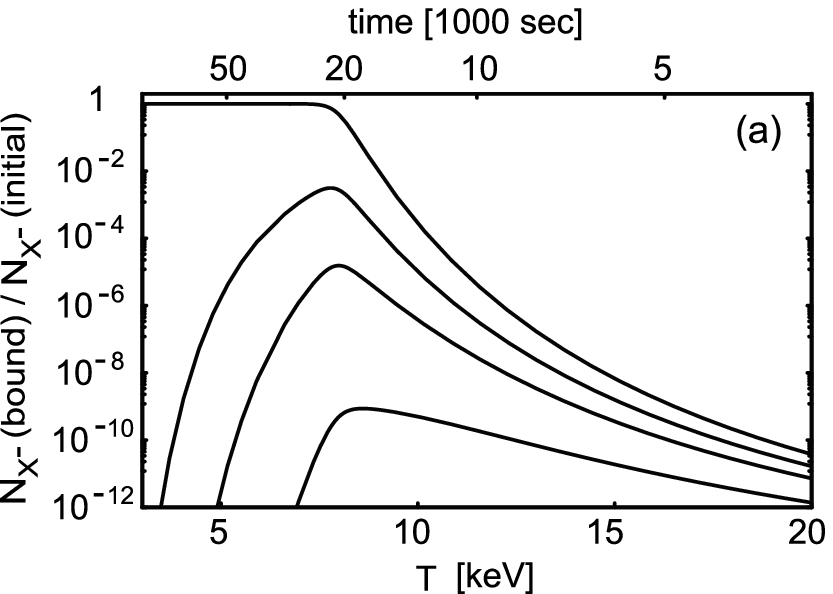,width=8cm}
\hspace{0.2cm}
\epsfig{file=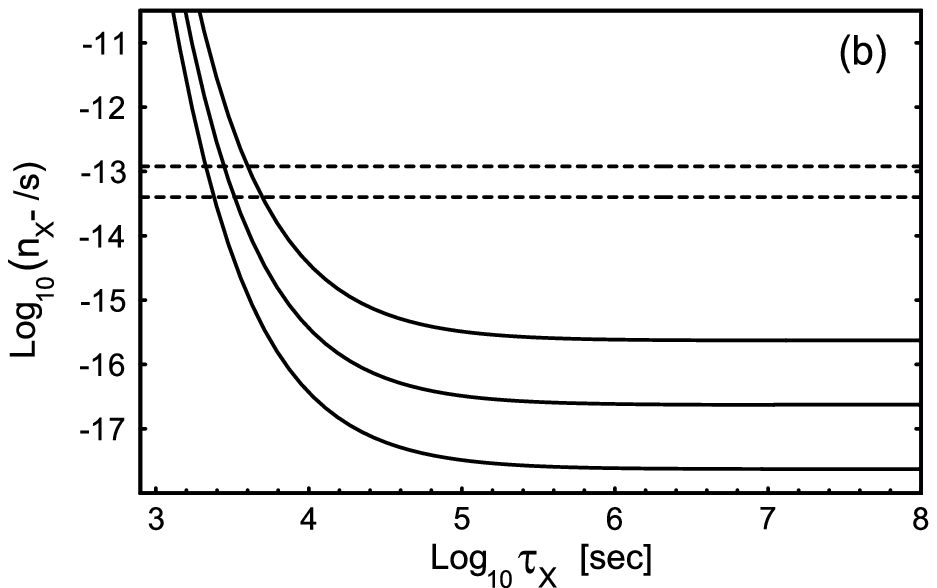,width=8cm}
\end{center}
\caption[]{(a) The number 
fraction of $X^-$ particles which form bound states with $^4 {\rm He}$ 
as a function of temperature $T$. $\tau_X = (1, 2, 4, \infty)\times 10^3$ sec from
 bottom to top. (b)
The contour plot of $^6 {\rm Li}$ abundance produced by the CBBN process in 
$(\tau_X,\, n_{X^-}/s)$ plane. The solid lines represent $^6 {\rm Li}=10^{-12}$, $10^{-11}$, and $10^{-10}$ from the bottom to the top. The dashed lines represent the thermal relic abundance of stau for $m_{\widetilde{\tau}}=100$ GeV (below) and 300 GeV (above).}
\label{fig:stau_fig6}
\end{figure}

 Before closing, let us briefly discuss the particle physics implication of our result.
With the cross section $\langle \, \sigma v \,\rangle$ in 
Eq.~(\ref{eq:sigmav}), the $^6{\rm Li}$ production via CBBN is described by
\begin{equation}
\left.
\frac{d}{dt}\, ^6{\rm Li}
\right|_{\rm CBBN}
 \;=\;
n_{\rm BS}\langle \, \sigma v \,\rangle {\rm D} ,
\label{eq:ddtLi6}
\end{equation}
where
${\rm D}\equiv n_d/n_B$ and 
$^6{\rm Li}\equiv n_{^6 \rm Li}/n_B$ are the abundances of these elements normalized by the baryon number density. 
The ratio of the number density of the bound states $n_{\rm BS}$ 
to the entropy density is given
 by~\footnote{We have assumed that the reaction 
 $^4 {\rm He} + X^- \leftrightarrow (^4 {\rm He} X^-)$ is in chemical equilibrium. 
 For more detailed analysis of the evolution of the bound state abundance, 
 see Ref.~\cite{KT}.}
\begin{equation}
\frac{n_{\rm BS}}{s} \;=\; 
\left.\frac{n_{X^-}}{s}\right|_{\rm initial} 
\cdot
\frac{\exp(-{t/\tau_X})}
{1+ n_\alpha^{-1}(m_\alpha T / 2\pi)^{3/2} \exp (-E_b/T)} ,
\end{equation}
where $n_{\alpha}$ and $m_{\alpha}$ are the number density and the mass of the $^4 {\rm He}$, and $E_b$ GeV is the binding energy of $(^4{\rm He}X^-)$. 
The fraction of $X^-$  which form bound states with $^4 {\rm He}$ is shown 
in Fig.~\ref{fig:stau_fig6}(a), as a function of temperature $T$ for various lifetimes of $X$.\footnote{For our numerical calculation, we use
 $Y = 4n_\alpha/n_B = 0.25$, ${\rm D}=2.78\times 10^{-5}$, $n_B/s=0.87\times 10^{-10}$, 
 $s=1.715T^3$, $m_\alpha=3.73$ GeV \cite{RPP} and $E_b=337.33$ GeV (see the previous section).}
The bound state abundance is peaked at around $T=10$ keV for 
$\tau_{_X} > 1000$ sec,\footnote{For $\tau_{_X} < 1000$ sec the CBBN effect is negligible.
 See Fig.~\ref{fig:stau_fig6}(b).} which justifies the expansion in Eq.(\ref{eq:expansion}).
 Thus, the CBBN occurs at $t\gg 1000$ sec, when the SBBN processes ($t < 1000$ sec) are 
 already frozen. At such late time, the effect of the dissociation  
 $^6 {\rm Li}+p\to ^3{\rm He} + ^6{\rm He}$ is also negligible~\cite{KKM}, 
 and hence we have neglected it in Eq.~(\ref{eq:ddtLi6}).

Fig.~\ref{fig:stau_fig6}(b) shows the contour plot of $^6 {\rm Li}$ abundance produced by the CBBN process, as a function of initial $X^-$ abundance $n_{X^-}/s$ 
and the lifetime $\tau_{_X}$. 
In the limit of long lifetime, one obtains
\begin{equation}
\left.^6 {\rm Li}\right|_{\rm CBBN}\;\simeq\; 4.3\times 10^{-11}
\left(\frac{n_{X^-}/s}{10^{-16}}\right)
\left(\frac{\rm D}{2.78\times 10^{-5}}\right)\;.
\end{equation}
Therefore, the observational upper bound on the $^6 {\rm Li}$ abundance 
$^6 {\rm Li}<6.1\times 10^{-11}$ ($2\sigma$)~\cite{KKM} leads to a bound on the $X^-$ abundance, $n_{X^-}/s<1.4\times 10^{-16}$.
On the other hand, the thermal relic abundance of stau is given by
$n_{\widetilde{\tau}^-}/s  \simeq (4-8)\times 10^{-14}
(m_{\widetilde{\tau}} / 100~\mathrm{GeV})$~\cite{Ystau}. (Conservative cases are shown 
in Fig.~\ref{fig:stau_fig6}(b) as the dashed lines for 
 $m_{\widetilde{\tau}}=100$ GeV and 300 GeV). 
 Therefore, in the limit of long life time, an entropy production 
 with a dilution factor
 $\Delta \simeq (300-600)\times (m_{\widetilde{\tau}} / 100~\mathrm{GeV})$ 
 for the primordial stau abundance is necessary to avoid the overproduction of $^6{\rm Li}$.\footnote{With this dilution factor, the masses of stau and gravitino considered here become also consistent with other BBN constraints~\cite{BHIY}.}
 
On the other hand, this means that the primordial baryon asymmetry is also diluted 
by $\Delta$. Standard thermal leptogenesis~\cite{FY}, an attractive mechanism for 
baryogenesis, can normally explain the observed baryon asymmetry if the reheating
temperature satisfies $T_R\gsim 2\times 10^9$ GeV~\cite{Buchmuller:2004nz}. 
Since the maximal baryon 
asymmetry is proportional to $T_R$ in thermal leptogenesis, this means that,
for a late time dilution $\Delta \simeq (300-600)\times (m_{\widetilde{\tau}} / 100~\mathrm{GeV})$, the
reheating temperature of $T_R \gsim (0.6-1.2)\times 10^{12}~{\rm GeV}\times (m_{\widetilde{\tau}} / 100~\mathrm{GeV})$ is necessary. 
Such a reheating temperature can  be obtained in inflationary models.

\begin{acknowledgments}
  T.H., M.K. and Y.K. were
   partly supported by the Grant-in-Aid of MEXT No. 18540253.
  Useful discussions with Emiko Hiyama are gratefully acknowledged. 
\end{acknowledgments}


\end{document}